# Mathematical Definition, Mapping, and Detection of (Anti)Fragility

**N. N. Taleb**
NYU-Poly

**R Douady**
CNRS- Risk Data

**First Version, August 2012**

## Abstract

We provide a mathematical definition of fragility and antifragility as negative or positive sensitivity to a semi-measure of dispersion and volatility (a variant of negative or positive "vega") and examine the link to nonlinear effects. We integrate model error (and biases) into the fragile or antifragile context. Unlike risk, which is linked to psychological notions such as subjective preferences (hence cannot apply to a coffee cup) we offer a measure that is universal and concerns any object that has a probability distribution (whether such distribution is known or, critically, unknown).

We propose a detection of fragility, robustness, and antifragility using a single "fast-and-frugal", model-free, probability free heuristic that also picks up exposure to model error. The heuristic lends itself to immediate implementation, and uncovers hidden risks related to company size, forecasting problems, and bank tail exposures (it explains the forecasting biases). While simple to implement, it outperforms stress testing and other such methods such as Value-at-Risk.

# What is Fragility?

The notions of *fragility* and *antifragility* were introduced in Taleb(2011,2012). In short, *fragility* is related to how a system suffers from the variability of its environment beyond a certain preset threshold (when threshold is $K$, it is called $K$-fragility), while *antifragility* refers to when it benefits from this variability —in a similar way to "vega" of an option or a nonlinear payoff, that is, its sensitivity to volatility or some similar measure of scale of a distribution.

Simply, a coffee cup on a table suffers more from large deviations than from the cumulative effect of some shocks—conditional on being unbroken, it has to suffer more from "tail" events than regular ones around the center of the distribution, the "at the money" category. This is the case of elements of nature that have survived: conditional on being in existence, then the class of events around the mean should matter considerably less than tail events, particularly when the probabilities decline faster than the inverse of the harm, which is the case of all used monomodal probability distributions. Further, what has exposure to tail events suffers from uncertainty; typically, when systems – a building, a bridge, a nuclear plant, an airplane, or a bank balance sheet– are made robust to a certain level of variability and stress but may fail or collapse if this level is exceeded, then they are particularly *fragile* to uncertainty about the distribution of the stressor, hence to model error, as this uncertainty increases the probability of dipping below the robustness level, bringing a higher probability of collapse. In the opposite case, the natural selection of an evolutionary process is particularly *antifragile*, indeed, a more volatile environment increases the survival rate of robust species and eliminates those whose superiority over other species is highly dependent on environmental parameters.

Figure 1 show the "tail vega" sensitivity of an object calculated discretely at two different lower absolute mean deviations. We use for the purpose of fragility and antifragility, in place of measures in $L^2$ such as standard deviations, which restrict the choice of probability distributions, the broader measure of absolute deviation, cut into two parts: lower and upper semi-deviation above the distribution center $\Omega$.

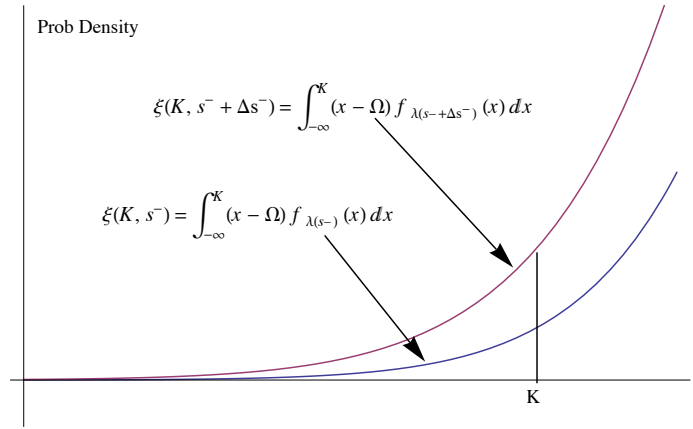

*Figure 1- A definition of fragility as left tail-vega sensitivity; the figure shows the effect of the perturbation of the lower semi-deviation $s^-$ on the tail integral $\xi$ of $(x - \Omega)$ below K, with $\Omega$ entering constant. Our detection of fragility does not require the specification of f the probability distribution.*

This article aims at providing a proper mathematical definition of fragility, robustness, and antifragility and examining how these apply to different cases where this notion is applicable.

**Intrinsic and Inherited Fragility:** Our definition of fragility is two-fold. First, of concern is the intrinsic fragility, the shape of the probability distribution of a variable and its sensitivity to $s^-$, a parameter controlling the left side of its own distribution. But we do not often directly observe the statistical distribution of objects, and, if we did, it would be difficult to measure their tail-vega sensitivity. Nor do we need to specify such distribution: we can gauge the response of a given object to the volatility of an external stressor that affects it. For instance, an option is usually analyzed with respect to the scale of the distribution of the "underlying" security, not its own; the fragility of a coffee cup is determined as a response to a given source of randomness or stress; that of a house with respect of, among other sources, the distribution of earthquakes. This fragility coming from the effect of the underlying is called inherited fragility. The transfer function, which we present next, allows us to assess the effect, increase or decrease in fragility, coming from changes in the underlying source of stress.

**Transfer Function:** A nonlinear exposure to a certain source of randomness maps into tail-vega sensitivity (hence fragility). We prove that

<div style="text-align:center">Inherited Fragility $\Leftrightarrow$ Concavity in exposure on the left side of the distribution</div>

and build $H$, a transfer function giving an exact mapping of tail vega sensitivity to the second derivative of a function. The transfer function will allow us to probe parts of the distribution and generate a fragility-detection heuristic covering both physical fragility and model error.

### *FRAGILITY AS SEPARATE RISK FROM PSYCHOLOGICAL PREFERENCES*

**Avoidance of the Psychological**: We start from the definition of fragility as tail vega sensitivity, and end up with nonlinearity as a necessary attribute of the source of such fragility in the inherited case —a cause of the disease rather than the disease itself. However, there is a long literature by economists and decision scientists embedding risk into psychological preferences —historically, risk has been described as derived from risk aversion as a result of the structure of choices under uncertainty with a concavity of the muddled concept of "utility" of payoff, see Pratt (1964), Arrow (1965), Rothchild and Stiglitz(1970,1971). But this "utility" business never led anywhere except the circularity, expressed by Machina and Rothschild (2008), "risk is what risk-averters hate." Indeed limiting risk to aversion to concavity of choices is a quite unhappy result —the utility curve cannot be possibly monotone concave, but rather, like everything in nature necessarily bounded on both sides, the left and the right, convex-concave and, as Kahneman and Tversky (1979) have debunked, both path dependent and mixed in its nonlinearity.

**Beyond Jensen's Inequality**: Furthermore, the economics and decision-theory literature reposes on the effect of Jensen's inequality, an analysis which requires monotone convex or concave transformations —in fact limited to the expectation operator. The world is unfortunately more complicated in its nonlinearities. Thanks to the transfer function we can accommodate situations where the source is not merely convex, but convex-concave and any other form of mixed nonlinearities common in exposures, which includes nonlinear dose-response in biology. For instance, the application of the transfer function to the Kahneman-Tversky value function, convex in the negative domain and concave in the positive one, shows that its decreases fragility in the left tail (hence more robustness) and reduces the effect of the right tail as well (also more robustness), which allows to assert that we are psychologically "more robust" to changes in wealth than implied from the distribution of such wealth, which happens to be extremely fat-tailed.

Accordingly, our approach relies on nonlinearity of exposure as detection of the vega-sensitivity, not as a definition of fragility. And nonlinearity in a source of stress is necessarily associated with fragility. Clearly, a coffee cup, a house or a bridge don't have psychological preferences, subjective utility, etc. Yet they are concave in their reaction to harm: simply, taking $z$ as a stress level and $\Pi(z)$ the harm function, it suffices to see that, with $n > 1$,

$$\Pi(nz) < n \ \Pi(\ z) \quad \text{for all} \ \ 0 < nz < Z^*$$

where $Z^*$ is the level (not necessarily specified) at which the item is broken. Such inequality leads to $\Pi(z)$ having a negative second derivative at the initial value $z$.

So if a coffee cup is less harmed by $n$ times a stressor of intensity $Z$ than once a stressor of $nZ$, then harm (as a negative function) needs to be concave to stressors up to the point of breaking; such stricture is imposed by the structure of survival probabilities and the distribution of harmful events, and has nothing to do with subjective utility or some other figments. Just as with a large stone hurting more than the equivalent weight in pebbles, if, for a human, jumping one millimeter caused an exact linear fraction of the damage of, say, jumping to the ground from thirty feet, then the person would be already dead from cumulative harm. Actually a simple computation shows that he would have expired within hours from touching objects or pacing in his living room, given the multitude of such stressors and their total effect. The fragility that comes from linearity is immediately visible, so we rule it out because the object would be already broken and the person already dead. The relative frequency of ordinary events compared to extreme events is the determinant. In the financial markets, there are at least ten thousand times more events of 0.1% deviations than events of 10%. There are close to 8,000 micro-earthquakes daily on planet earth, that is, those below 2 on the Richter scale —about 3 million a year. These are totally harmless, and, with 3 million per year, you would need them to be so. But shocks of intensity 6 and higher on the scale make the newspapers. Accordingly, we are necessarily immune to the *cumulative* effect of small deviations, or shocks of very small magnitude, which implies that these affect us disproportionally less (that is, nonlinearly less) than larger ones.

Model error is not necessarily mean preserving. $s^-$, the lower absolute semi-deviation does not just express changes in overall dispersion in the distribution, such as for instance the "scaling" case, but also changes in the mean, i.e. when the upper semi-deviation from $\Omega$ to infinity is invariant, or even decline in a compensatory manner to make the overall mean absolute deviation unchanged. This would be the case when we shift the distribution instead of rescaling it. Thus the same vega-sensitivity can also express sensitivity to a stressor (dose increase) in medicine or other fields in its effect on either tail. Thus $s^-(\lambda)$ will allow us to express the sensitivity to the "disorder cluster" (Taleb, 2012): i) uncertainty, ii) variability, iii) imperfect, incomplete knowledge, iv) chance, v) chaos, vi) volatility, vii) disorder, viii) entropy, ix) time, x) the unknown, xi) randomness, xii) turmoil, xiii) stressor, xiv) error, xv) dispersion of outcomes.

### DETECTION HEURISTIC

Finally, thanks to the transfer function, this paper proposes a risk heuristic that "works" in detecting fragility even if we use the wrong model/pricing method/probability distribution. The main idea is that *a wrong ruler will not measure the height of a child; but it can certainly tell us if he is growing*. Since risks in the tails map to nonlinearities (concavity of exposure), second order effects reveal fragility, particularly in the tails where they map to large tail exposures, as revealed through perturbation analysis. More generally every nonlinear function will produce some kind of positive or negative exposures to volatility for some parts of the distribution.

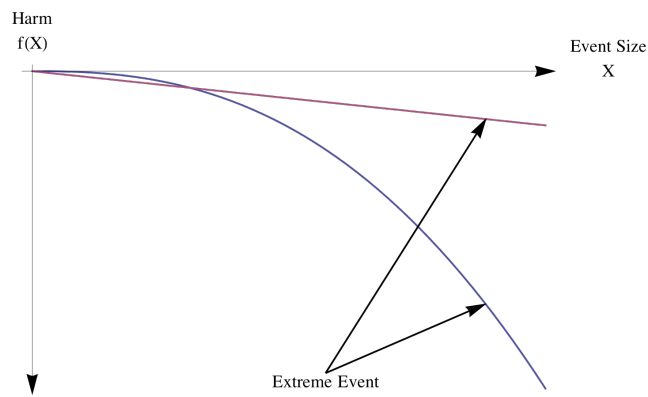

Figure 2- Disproportionate effect of tail events on nonlinear exposures, illustrating the necessity of nonlinearity of the harm function and showing how we can extrapolate outside the model to probe unseen fragility.

**Fragility and Model Error:** As we saw this definition of fragility extends to model error, as some models produce negative sensitivity to uncertainty, in addition to effects and biases under variability. So, beyond physical fragility, the same approach measures model fragility, based on the difference between a *point estimate* and stochastic value (i.e., full distribution). Increasing the variability (say, variance) of the estimated value (but not the mean), may lead to one-sided effect on the model —just as an increase of volatility causes porcelain cups to break. Hence sensitivity to the volatility of such value, the "vega" of the model with respect to such value is no different from the vega of other payoffs. For instance, the misuse of thin-tailed distributions (say Gaussian) appears immediately through perturbation of the standard deviation, no longer used as point estimate, but as a distribution with its own variance.  For instance, it can be shown how fat-tailed (e.g. power-law tailed) probability distributions can be expressed by simple nested perturbation and mixing of Gaussian ones. Such a representation pinpoints the fragility of a wrong probability model and its consequences in terms of underestimation of risks, stress tests and similar matters.

**Antifragility**: It is not quite the mirror image of fragility, as it implies positive vega above some threshold in the positive tail of the distribution and absence of fragility in the left tail, which leads to a distribution that is skewed right.

# Fragility and Transfer Theorems

Table 1- Introduces the Exhaustive Taxonomy of all Possible Payoffs y=f(x)

| Type | Condition | Left Tail (loss domain) | Right Tail (gains domain) | Nonlinear Payoff Function $y = f(x)$ "derivative", where $x$ is a random variable | Derivatives Equivalent (Taleb, 1997) | Effect of fataledness of f(x) compared to primitive $x$ |
|---|---|---|---|---|---|---|
| Type 1 | Fragile (type 1) | Fat (regular or absorbing barrier) | Fat | Mixed concave left, convex right (fence) | Long up −vega, short down − vega | More fragility if absorbing barrier, neutral otherwise |
| Type 2 | Fragile (type 2) | Fat | Thin | Concave | Short vega | More fragility |
| Type 3 | Robust | Thin | Thin | Mixed convex left, concave right (digital, sigmoid) | Short down − vega, long up − vega | No effect |
| Type 4 | Antifragile | Thin | Fat (thicker than left) | Convex | Long vega | More antifragility |

The central Table 1 introduces the exhaustive map of possible outcomes, with 4 mutually exclusive categories of payoffs.

Our steps in the rest of the paper are as follows:

**a.** We provide a mathematical definition of fragility, robustness and antifragility.

**b.** We present the problem of measuring tail risks and show the presence of severe biases attending the estimation of small probability and its nonlinearity (convexity) to parametric (and other) perturbations.

**c.** We express the concept of model fragility in terms of left tail exposure, and show correspondence to the concavity of the payoff from a random variable.

**d.** Finally, we present our simple heuristic to detect the possibility of both fragility and model error across a broad range of probabilistic estimations.

Conceptually, *fragility* resides in the fact that a small – or at least reasonable – uncertainty on the macro-parameter of a distribution may have dramatic consequences on the result of a given stress test, or on some measure that depends on the left tail of the distribution, such as an out-of-the-money option. This hypersensitivity of what we like to call an "out of the money put price" to the macro-parameter, which is *some* measure of the volatility of the distribution of the underlying source of randomness.

Formally, fragility is defined as the sensitivity of the left-tail shortfall (non-conditioned by probability) below a certain threshold K to the overall left semi-deviation of the distribution.

*Examples*

**a.** Example: a porcelain coffee cup subjected to  random daily stressors from use.

**b.** Example: tail distribution in the function of the arrival time of an aircraft.

**c.** Example: hidden risks of famine to a population subjected to monoculture —or, more generally, fragilizing errors in the application of Ricardo's comparative advantage without taking into account second order effects.

**d.** Example: hidden tail exposures to budget deficits' nonlinearities to unemployment.

**e.** Example: hidden tail exposure from dependence on a source of energy, etc. ("squeezability argument").\

### Tail Vega Sensitivity

We construct a measure of "vega" in the tails of the distribution that depends on the variations of $s$, the semi-deviation below a certain level $\Omega$, chosen in the $L^1$ norm in order to insure its existence under "fat tailed" distributions with finite first semi-moment. In fact $s$ would exist as a measure even in the case of infinite moments to the right side of $\Omega$.

Let $X$ be a random variable, the distribution of which is one among a one-parameter family of pdf $f_\lambda$, $\lambda \in I \subset \mathbb{R}$. We consider a fixed reference value $\Omega$ and, from this reference, the left-semi-absolute deviation:

$$s^-(\lambda) = \int_{-\infty}^{\Omega} (\Omega - x) f_\lambda(x) dx$$

We assume that $\lambda \to s^-(\lambda)$ is continuous, strictly increasing and spans the whole range $\mathbb{R}_+ = [0, +\infty)$, so that we may use the left-semi-absolute deviation $s^-$ as a parameter by considering the inverse function $\lambda(s): \mathbb{R}_+ \to I$, defined by $s^-(\lambda(s)) = s$ for $s \in \mathbb{R}_+$.

This condition is for instance satisfied if, for any given $x < \Omega$, the probability is a continuous and increasing function of $\lambda$. Indeed, denoting $F_\lambda(x) = \Pr_{f_\lambda}(X < x) = \int_{-\infty}^{x} f_\lambda(t) dt$, an integration by part yields:

$$s^-(\lambda) = \int_{-\infty}^{\Omega} F_\lambda(x) dx$$

This is the case when $\lambda$ is a scaling parameter, i.e. $X \sim \Omega + \lambda(X_1 - \Omega)$, indeed one has in this case $F_\lambda(x) = F_1\left(\Omega + \dfrac{x - \Omega}{\lambda}\right)$, $\dfrac{\partial F_\lambda}{\partial \lambda}(x) = \dfrac{\Omega - x}{\lambda} f_\lambda(x)$ and $s^-(\lambda) = \lambda s^-(1)$.

It is also the case when $\lambda$ is a shifting parameter, i.e. $X \sim X_0 - \lambda$, indeed, in this case $F_\lambda(x) = F_0(x + \lambda)$ and $\dfrac{\partial s^-}{\partial \lambda} = F_\lambda(\Omega)$ .

For $K < \Omega$ and $s \in \mathbb{R}_+$, let:

$$\xi(K, s^-) = \int_{-\infty}^{K} (\Omega - x) f_{\lambda(s^-)}(x) dx$$

In particular, $\xi(\Omega, s^-) = s^-$. We assume, in a first step, that the function $\xi(K, s^-)$ is differentiable on $(-\infty, \Omega] \times \mathbb{R}_+$. The *K-left-tail-vega sensitivity* of $X$ at stress level $K < \Omega$ and deviation level $s^- > 0$ for the pdf $f_\lambda$ is:

$$V(X, f_\lambda, K, s^-) = \frac{\partial \xi}{\partial s^-}(K, s^-) = \left(\int_{-\infty}^{K} (\Omega - x) \frac{\partial f_\lambda}{\partial \lambda}(x) dx\right)\left(\frac{ds^-}{d\lambda}\right)^{-1}$$

As the in many practical instances where threshold effects are involved, it may occur that $\xi$ does not depend smoothly on $s^-$. We therefore also define a *finite difference* version of the *vega-sensitivity* as follows:

$$V(X, f_\lambda, K, s^-, \Delta s) = \frac{1}{2\Delta s}\Big(\xi(K, s^- + \Delta s) - \xi(K, s^- - \Delta s)\Big)$$

$$= \int_{-\infty}^{K} (\Omega - x) \frac{f_{\lambda(s^- + \Delta s)}(x) - f_{\lambda(s^- - \Delta s)}(x)}{2\Delta s} dx$$

Hence omitting the input $\Delta s$ implicitly assumes that $\Delta s \to 0$.

Note that $\xi\left(K, s^-\right) = -\mathrm{E}_{f_\lambda}\left[X \,\middle|\, X < K\right]\mathrm{Pr}_{f_\lambda}\left(X < K\right)$. It can be decomposed into two parts:

$$\xi\left(K, s^-(\lambda)\right) = (\Omega - K)F_\lambda(K) + P_\lambda(K)$$

$$P_\lambda(K) = \int_{-\infty}^{K}(K - x)f_\lambda(x)dx$$

Where the first part $(\Omega - K)F_\lambda(K)$ is proportional to the probability of the variable being below the stress level $K$ and the second part $P_\lambda(K)$ is the expectation of the amount by which $X$ is below $K$ (counting 0 when it is not). Making a parallel with financial options, while $s^-(\lambda)$ is a "put at-the-money", $\xi(K, s^-)$ is the sum of a put struck at $K$ and a digital put also struck at $K$ with amount $\Omega - K$; it can equivalently be seen as a put struck at $\Omega$ with a down-and-in European barrier at $K$.

Letting $\lambda = \lambda(s^-)$ and integrating by part yields

$$\xi\left(K, s^-(\lambda)\right) = (\Omega - K)F_\lambda(K) + \int_{-\infty}^{K}F_\lambda(x)dx = \int_{-\infty}^{\Omega}F_\lambda^K(x)dx$$

Where $F_\lambda^K(x) = F_\lambda\left(\min(x, K)\right) = \min\left(F_\lambda(x), F_\lambda(K)\right)$, so that

$$V(X, f_\lambda, K, s^-) = \frac{\partial \xi}{\partial s}(K, s^-) = \frac{\int_{-\infty}^{\Omega} \dfrac{\partial F_\lambda^K}{\partial \lambda}(x)dx}{\int_{-\infty}^{\Omega} \dfrac{\partial F_\lambda}{\partial \lambda}(x)dx}$$

For finite differences

$$V(X, f_\lambda, K, s^-, \Delta s) = \frac{1}{2\Delta s}\int_{-\infty}^{\Omega}\left(\Delta F_{\lambda,\Delta s}^K(x)\right)dx$$

Where $\lambda_s^+$ and $\lambda_s^-$ are such that $s(\lambda_{s^-}^+) = s^- + \Delta s$, $s(\lambda_{s^-}^-) = s^- - \Delta s$ and $\Delta F_{\lambda,\Delta s}^K(x) = F_{\lambda_s^+}^K(x) - F_{\lambda_s^-}^K(x)$.

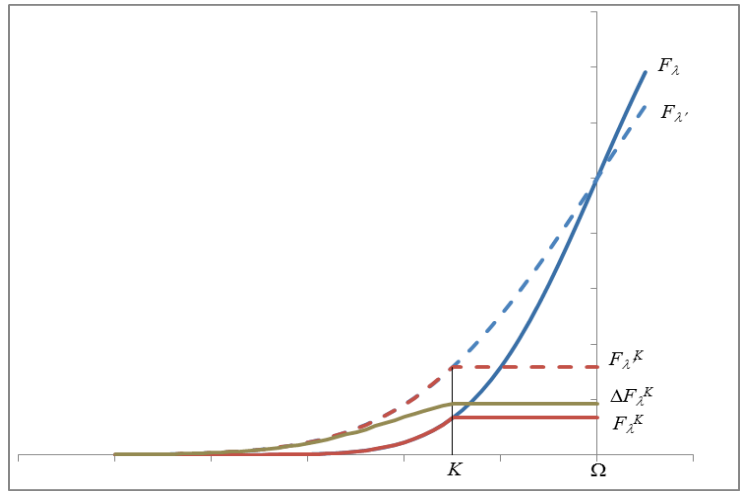

*Figure 3- The different curves of $F_\lambda(K)$ and $F'_\lambda(K)$ showing the difference in sensitivity to changes in at different levels of K.*

### Mathematical Expression of Fragility

In essence, fragility is the sensitivity of a given risk measure to an error in the estimation of the (possibly one-sided) deviation parameter of a distribution, especially due to the fact that the risk measure involves parts of the distribution – tails – that are away from the portion used for estimation. The risk measure then assumes certain extrapolation rules that have first order consequences. These consequences are even more amplified when the risk measure applies to a variable that is derived from that used for estimation, when the relation between the two variables is strongly nonlinear, as is often the case.

## Definition of Fragility: The *Intrinsic* Case

The <u>local fragility</u> of a random variable $X_\lambda$ depending on parameter $\lambda$, at stress level $K$ and semi-deviation level $s^-(\lambda)$ with pdf $f_\lambda$ is its $K$-left-tailed semi-vega sensitivity $V(X, f_\lambda, K, s^-)$.

The <u>finite-difference fragility</u> of $X_\lambda$ at stress level $K$ and semi-deviation level $s^-(\lambda) \pm \Delta s$ with pdf $f_\lambda$ is its $K$-left-tailed finite-difference semi-vega sensitivity $V(X, f_\lambda, K, s^-, \Delta s)$.

In this definition, the *fragility* relies in the unsaid assumptions made when extrapolating the distribution of $X_\lambda$ from areas used to estimate the semi-absolute deviation $s^-(\lambda)$, around $\Omega$, to areas around $K$ on which the risk measure $\xi$ depends.

## Definition of Fragility: The *Inherited* Case

We here consider the particular case where a random variable $Y = \varphi(X)$ depends on another source of risk $X$, itself subject to a parameter $\lambda$. Let us keep the above notations for $X$, while we denote by $g_\lambda$ the pdf of $Y$, $\Omega_Y = \varphi(\Omega)$ and $u^-(\lambda)$ the left-semi-deviation of $Y$. Given a "strike" level $L = \varphi(K)$, let us define, as in the case of $X$ :

$$\zeta(L, u^-(\lambda)) = \int_{-\infty}^{K} (\Omega_Y - y) g_\lambda(y) dy$$

The <u>inherited fragility</u> of $Y$ with respect to $X$ at stress level $L = \varphi(K)$ and left-semi-deviation level $s^-(\lambda)$ of $X$ is the partial derivative:

$$V_X(Y, g_\lambda, L, s^-(\lambda)) = \frac{\partial \zeta}{\partial s}(L, u^-(\lambda)) = \left( \int_{-\infty}^{K} (\Omega_Y - y) \frac{\partial g_\lambda}{\partial \lambda}(y) dy \right) \left( \frac{ds^-}{d\lambda} \right)^{-1}$$

Note that the stress level and the pdf are defined for the variable $Y$, but the parameter which is used for differentiation is the left-semi-absolute deviation of $X$, $s^-(\lambda)$. Indeed, in this process, one first measures the distribution of $X$ and its left-semi-absolute deviation, then the function $\varphi$ is applied, using some mathematical model of $Y$ with respect to $X$ and the risk measure $\zeta$ is estimated. If an error is made when measuring $s^-(\lambda)$, its impact on the risk measure of $Y$ is amplified by the ratio given by the "inherited fragility".

Once again, one may use finite differences and define the <u>finite-difference inherited fragility</u> of $Y$ with respect to $X$, by replacing, in the above equation, differentiation by finite differences between values $\lambda^+$ and $\lambda^-$, where $s^-(\lambda^+) = s^- + \Delta s$ and $s^-(\lambda^-) = s^- - \Delta s$.

## Implications of a Nonlinear Change of Variable on the Intrinsic Fragility

We here study the case of a random variable $Y = \varphi(X)$, the pdf $g_\lambda$ of which also depends on parameter $\lambda$, related to a variable $X$ by the nonlinear function $\varphi$. We are now interested in comparing their *intrinsic fragilities*. We shall say, for instance, that $Y$ is *more fragile* at the stress level $L$ and left-semi-deviation level $u^-(\lambda)$ than the random variable $X$, at stress level $K$ and left-semi-deviation level $s^-(\lambda)$ if the $L$-left-tailed semi-vega sensitivity of $Y_\lambda$ is higher than the $K$-left-tailed semi-vega sensitivity of $X_\lambda$:

$$V(Y, g_\lambda, L, u^-) > V(X, f_\lambda, K, s^-)$$

One may use finite differences to compare the fragility of two random variables: $V(Y, g_\lambda, L, u^-, \Delta u) > V(X, f_\lambda, K, s^-, \Delta s)$. In this case, finite variations must be comparable in size, namely $\Delta u / u^- = \Delta s / s^-$.

Let us assume, to start with, that $\varphi$ is differentiable, strictly increasing and scaled so that $\Omega_Y = \varphi(\Omega) = \Omega$. We also assume that, for any given $x < \Omega$, $\frac{\partial F_\lambda}{\partial \lambda}(x) > 0$ . In this case, as observed above, $\lambda \to s^-(\lambda)$ is also increasing.

Let us denote $G_\lambda(y) = \mathrm{Pr}_{g_\lambda}(Y < y)$ . We have:

$$G_\lambda\big(\varphi(x)\big) = \mathrm{Pr}_{g_\lambda}(Y < \varphi(x)) = \mathrm{Pr}_{f_\lambda}(X < x) = F_\lambda(x)$$

Hence, if $\zeta(L, u^-)$ denotes the equivalent of $\xi(K, s^-)$ with variable $(Y, g_\lambda)$ instead of $(X, f_\lambda)$, then we have:

$$\zeta(L, u^-(\lambda)) = \int_{-\infty}^{\Omega} G_\lambda^L(y)\, dy = \int_{-\infty}^{\Omega} F_\lambda^K(x) \frac{d\varphi}{dx}(x)\, dx$$

Because $\varphi$ is increasing and $\min(\varphi(x), \varphi(K)) = \varphi(\min(x, K))$. In particular

$$u^-(\lambda) = \zeta(\Omega, u^-(\lambda)) = \int_{-\infty}^{\Omega} F_\lambda(x) \frac{d\varphi}{dx}(x)\, dx$$

The $L$-left-tail-vega sensitivity of $Y$ is therefore:

$$V(Y, g_\lambda, L, u^-(\lambda)) = \frac{\int_{-\infty}^{\Omega} \frac{\partial F_\lambda^K}{\partial \lambda}(x) \frac{d\varphi}{dx}(x)\, dx}{\int_{-\infty}^{\Omega} \frac{\partial F_\lambda}{\partial \lambda}(x) \frac{d\varphi}{dx}(x)\, dx}$$

For finite variations:

$$V(Y, g_\lambda, L, u^-(\lambda), \Delta u) = \frac{1}{2\Delta u} \int_{-\infty}^{\Omega} \Delta F_{\lambda, \Delta u}^K(x) \frac{d\varphi}{dx}(x)\, dx$$

Where $\lambda_{u^-}^+$ and $\lambda_{u^-}^-$ are such that $u(\lambda_{u^-}^+) = u^- + \Delta u$ , $u(\lambda_{u^-}^-) = u^- - \Delta u$ and $F_{\lambda, \Delta u}^K(x) = F_{\lambda_{u^-}^+}^K(x) - F_{\lambda_{u^-}^-}^K(x)$ .

Next, Theorem 1 proves how a concave transformation $\varphi(x)$ of a random variable $x$ produces fragility.

### *Theorem 1 (Fragility Transfer Theorem)*

*Let, with the above notations, $\varphi : \mathbb{R} \to \mathbb{R}$ be a twice differentiable function such that $\varphi(\Omega) = \Omega$ and for any $x < \Omega$, $\frac{d\varphi}{dx}(x) > 0$ . The random variable $Y = \varphi(X)$ is more fragile at level $L = \varphi(K)$ and pdf $g_\lambda$ than $X$ at level $K$ and pdf $f_\lambda$ if, and only if, one has:*

$$\int_{-\infty}^{\Omega} H_\lambda^K(x) \frac{d^2\varphi}{dx^2}(x)\, dx < 0$$

*Where*

$$H_\lambda^K(x) = \frac{\partial P_\lambda^K}{\partial \lambda}(x) \Big/ \frac{\partial P_\lambda^K}{\partial \lambda}(\Omega) - \frac{\partial P_\lambda}{\partial \lambda}(x) \Big/ \frac{\partial P_\lambda}{\partial \lambda}(\Omega)$$

*and where $P_\lambda(x) = \int_{-\infty}^{x} F_\lambda(t)\, dt$ is the price of the "put option" on $X_\lambda$ with "strike" $x$ and $P_\lambda^K(x) = \int_{-\infty}^{x} F_\lambda^K(t)\, dt$ is that of a "put option" with "strike" $x$ and "European down-and-in barrier" at $K$.*

$H$ can be seen as a *transfer function*, expressed as the difference between two ratios. For a given level $x$ of the random variable on the left hand side of $\Omega$, the second one is the ratio of the vega of a put struck at $x$ normalized by that of a put "at the money" (i.e. struck at $\Omega$), while the first one is the same ratio, but where puts struck at $x$ and $\Omega$ are "European down-and-in options" with triggering barrier at the level $K$.

## Proof

Let $I_{X_\lambda} = \int_{-\infty}^{\Omega} \frac{\partial F_\lambda}{\partial \lambda}(x)\, dx$, $I_{X_\lambda}^K = \int_{-\infty}^{\Omega} \frac{\partial F_\lambda^K}{\partial \lambda}(x)\, dx$, $I_{Y_\lambda} = \int_{-\infty}^{\Omega} \frac{\partial F_\lambda}{\partial \lambda}(x) \frac{d\varphi}{dx}(x)\, dx$ and $I_{Y_\lambda}^L = \int_{-\infty}^{\Omega} \frac{\partial F_\lambda^K}{\partial \lambda}(x) \frac{d\varphi}{dx}(x)\, dx$ . One has $V(X, f_\lambda, K, s^-(\lambda)) = I_{X_\lambda}^K \big/ I_{X_\lambda}$ and $V(Y, g_\lambda, L, u^-(\lambda)) = I_{Y_\lambda}^L \big/ I_{Y_\lambda}$ hence:

$$V(Y, g_\lambda, L, u^-(\lambda)) - V(X, f_\lambda, K, s^-(\lambda)) = \frac{I_{Y_\lambda}^L}{I_{Y_\lambda}} - \frac{I_{X_\lambda}^K}{I_{X_\lambda}} = \frac{I_{X_\lambda}^K}{I_{Y_\lambda}} \left( \frac{I_{Y_\lambda}^L}{I_{X_\lambda}^K} - \frac{I_{Y_\lambda}}{I_{X_\lambda}} \right)$$

Therefore, because the four integrals are positive, $V(Y, g_\lambda, L, u^-(\lambda)) - V(X, f_\lambda, K, s^-(\lambda))$ has the same sign as

$I_{Y_\lambda}^L / I_{X_\lambda}^K - I_{Y_\lambda} / I_{X_\lambda}$. On the other hand, we have $I_{X_\lambda} = \frac{\partial P_\lambda}{\partial \lambda}(\Omega)$, $I_{X_\lambda}^K = \frac{\partial P_\lambda^K}{\partial \lambda}(\Omega)$ and

$$I_{Y_\lambda} = \int_{-\infty}^{\Omega} \frac{\partial F_\lambda}{\partial \lambda}(x) \frac{d\varphi}{dx}(x) dx = \frac{\partial P_\lambda}{\partial \lambda}(\Omega) \frac{d\varphi}{dx}(\Omega) - \int_{-\infty}^{\Omega} \frac{\partial P_\lambda}{\partial \lambda}(x) \frac{d^2\varphi}{dx^2}(x) dx$$

$$I_{Y_\lambda}^L = \int_{-\infty}^{\Omega} \frac{\partial F_\lambda^K}{\partial \lambda}(x) \frac{d\varphi}{dx}(x) dx = \frac{\partial P_\lambda^K}{\partial \lambda}(\Omega) \frac{d\varphi}{dx}(\Omega) - \int_{-\infty}^{\Omega} \frac{\partial P_\lambda^K}{\partial \lambda}(x) \frac{d^2\varphi}{dx^2}(x) dx$$

An elementary calculation yields:

$$\frac{I_{Y_\lambda}^L}{I_{X_\lambda}^K} - \frac{I_{Y_\lambda}}{I_{X_\lambda}} = -\left(\frac{\partial P_\lambda^K}{\partial \lambda}(\Omega)\right)^{-1} \int_{-\infty}^{\Omega} \frac{\partial P_\lambda^K}{\partial \lambda}(x) \frac{d^2\varphi}{dx^2} dx + \left(\frac{\partial P_\lambda}{\partial \lambda}(\Omega)\right)^{-1} \int_{-\infty}^{\Omega} \frac{\partial P_\lambda}{\partial \lambda}(x) \frac{d^2\varphi}{dx^2} dx$$

$$= -\int_{-\infty}^{\Omega} H_\lambda^K(x) \frac{d^2\varphi}{dx^2} dx$$

∎

Let us now examine the properties of the function $H_\lambda^K(x)$. For $x \le K$, we have $\frac{\partial P_\lambda^K}{\partial \lambda}(x) = \frac{\partial P_\lambda}{\partial \lambda}(x) > 0$ (the positivity is a consequence of that of

$\partial F_\lambda / \partial \lambda$), therefore $H_\lambda^K(x)$ has the same sign as $\frac{\partial P_\lambda}{\partial \lambda}(\Omega) - \frac{\partial P_\lambda^K}{\partial \lambda}(\Omega)$. As this is a strict inequality, it extends to an interval on the right hand side of

$K$, say $(-\infty, K']$ with $K < K' < \Omega$.

But on the other hand:

$$\frac{\partial P_\lambda}{\partial \lambda}(\Omega) - \frac{\partial P_\lambda^K}{\partial \lambda}(\Omega) = \int_K^{\Omega} \frac{\partial F_\lambda}{\partial \lambda}(x) dx - (\Omega - K) \frac{\partial F_\lambda}{\partial \lambda}(K)$$

For $K$ negative enough, $\frac{\partial F_\lambda}{\partial \lambda}(K)$ is smaller than its average value over the interval $[K, \Omega]$, hence $\frac{\partial P_\lambda}{\partial \lambda}(\Omega) - \frac{\partial P_\lambda^K}{\partial \lambda}(\Omega) > 0$.

We have proven the following theorem.

### Theorem 2 ( Fragility Exacerbation Theorem)

*With the above notations, there exists a threshold $\Theta_\lambda < \Omega$ such that, if $K \le \Theta_\lambda$ then $H_\lambda^K(x) > 0$ for $x \in (-\infty, \kappa_\lambda]$ with $K < \kappa_\lambda < \Omega$. As a consequence, if the change of variable $\varphi$ is concave on $(-\infty, \kappa_\lambda]$ and linear on $[\kappa_\lambda, \Omega]$, then Y is more fragile at $L = \varphi(K)$ than X at K.*

One can prove that, for a monomodal distribution, $\Theta_\lambda < \kappa_\lambda < \Omega$ (see discussion below), so whatever the stress level $K$ below the threshold $\Theta_\lambda$, it suffices that the change of variable $\varphi$ be concave on the interval $(-\infty, \Theta_\lambda]$ and linear on $[\Theta_\lambda, \Omega]$ for $Y$ to become more fragile at $L$ than $X$ at $K$. In practice, as long as the change of variable is concave around the stress level $K$ and has limited convexity/concavity away from $K$, the fragility of $Y$ is greater than that of $X$.

Figure 2 shows the shape of $H_\lambda^K(x)$ in the case of a Gaussian distribution where $\lambda$ is a simple scaling parameter ($\lambda$ is the standard deviation $\sigma$) and $\Omega = 0$. We represented $K = -2\lambda$ while in this Gaussian case, $\Theta_\lambda = -1.585\lambda$.

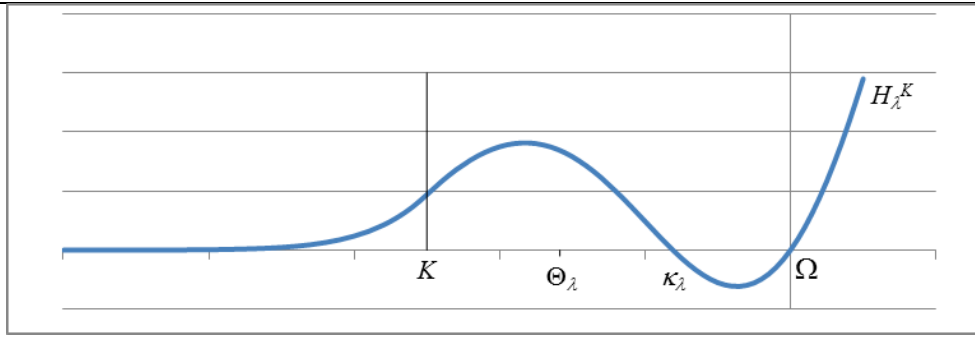

*Figure 4- The Transfer function H for different portions of the distribution: its sign flips in the region slightly below $\Omega$*

### DISCUSSION

### Monomodal case

We say that the family of distributions $(f_\lambda)$ is *left-monomodal* if there exists $\kappa_\lambda < \Omega$ such that $\dfrac{\partial f_\lambda}{\partial \lambda} \geq 0$ on $(-\infty, \kappa_\lambda]$ and $\dfrac{\partial f_\lambda}{\partial \lambda} \leq 0$ on $[\mu_\lambda, \Omega]$. In this case $\dfrac{\partial P_\lambda}{\partial \lambda}$ is a convex function on the left half-line $(-\infty, \mu_\lambda]$, then concave after the inflexion point $\mu_\lambda$. For $K \leq \mu_\lambda$, the function $\dfrac{\partial P_\lambda^K}{\partial \lambda}$ coincides with $\dfrac{\partial P_\lambda}{\partial \lambda}$ on $(-\infty, K]$, then is a linear extension, following the tangent to the graph of $\dfrac{\partial P_\lambda}{\partial \lambda}$ in $K$ (see graph below). The value of $\dfrac{\partial P_\lambda^K}{\partial \lambda}(\Omega)$ corresponds to the intersection point of this tangent with the vertical axis. It increases with $K$, from 0 when $K \to -\infty$ to a value above $\dfrac{\partial P_\lambda}{\partial \lambda}(\Omega)$ when $K = \mu_\lambda$. The threshold $\Theta_\lambda$ corresponds to the unique value of $K$ such that $\dfrac{\partial P_\lambda^K}{\partial \lambda}(\Omega) = \dfrac{\partial P_\lambda}{\partial \lambda}(\Omega)$. When $K < \Theta_\lambda$ then $G_\lambda(x) = \dfrac{\partial P_\lambda}{\partial \lambda}(x) \bigg/ \dfrac{\partial P_\lambda}{\partial \lambda}(\Omega)$ and $G_\lambda^K(x) = \dfrac{\partial P_\lambda^K}{\partial \lambda}(x) \bigg/ \dfrac{\partial P_\lambda^K}{\partial \lambda}(\Omega)$ are functions such that $G_\lambda(\Omega) = G_\lambda^K(\Omega) = 1$ and which are proportional for $x \leq K$, the latter being linear on $[K, \Omega]$. On the other hand, if $K < \Theta_\lambda$ then $\dfrac{\partial P_\lambda^K}{\partial \lambda}(\Omega) < \dfrac{\partial P_\lambda}{\partial \lambda}(\Omega)$ and $G_\lambda(K) < G_\lambda^K(K)$, which implies that $G_\lambda(x) < G_\lambda^K(x)$ for $x \leq K$. An elementary convexity analysis shows that, in this case, the equation $G_\lambda(x) = G_\lambda^K(x)$ has a unique solution $\kappa_\lambda$ with $\mu_\lambda < \kappa_\lambda < \Omega$. The "transfer" function $H_\lambda^K(x)$ is positive for $x < \kappa_\lambda$, in particular when $x \leq \mu_\lambda$ and negative for $\kappa_\lambda < x < \Omega$.

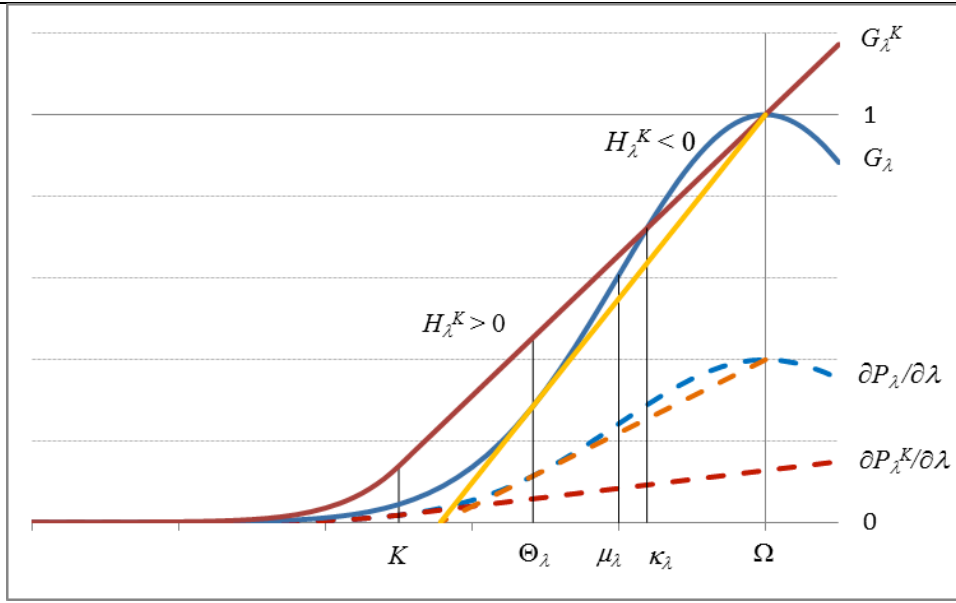

*Figure 5- The distribution $G_\lambda$ and the various derivatives of the unconditional shortfalls*

## Scaling Parameter

We assume here that $\lambda$ is a scaling parameter, i.e. $X_\lambda = \Omega + \lambda(X_1 - \Omega)$. In this case, as we saw above, we have $f_\lambda(x) = \dfrac{1}{\lambda} f_1\left(\Omega + \dfrac{x - \Omega}{\lambda}\right)$,

$F_\lambda(x) = F_1\left(\Omega + \dfrac{x - \Omega}{\lambda}\right)$, $P_\lambda(x) = \lambda P_1\left(\Omega + \dfrac{x - \Omega}{\lambda}\right)$ and $s^-(\lambda) = \lambda s^-(1)$. Hence

$$\xi(K, s^-(\lambda)) = (\Omega - K)F_1\left(\Omega + \dfrac{K - \Omega}{\lambda}\right) + \lambda P_1\left(\Omega + \dfrac{K - \Omega}{\lambda}\right)$$

$$\frac{\partial \xi}{\partial s^-}(K, s^-) = \frac{1}{s^-(1)} \frac{\partial \xi}{\partial \lambda}(K, \lambda) = \frac{1}{s^-(\lambda)}\left(P_\lambda(K) + (\Omega - K)F_\lambda(K) + (\Omega - K)^2 f_\lambda(K)\right)$$

When we apply a nonlinear transformation $\varphi$, the action of the parameter $\lambda$ is no longer a scaling: when small negative values of $X$ are multiplied by a scalar $\lambda$, so are large negative values of $X$. The scaling $\lambda$ applies to small negative values of the transformed variable $Y$ with a coefficient $\dfrac{d\varphi}{dx}(0)$, but large negative values are subject to a different coefficient $\dfrac{d\varphi}{dx}(K)$, which can potentially be very different.

## Fragility Drift

*Fragility* is defined at as the sensitivity – i.e. the first partial derivative – of the tail estimate $\xi$ with respect to the left semi-deviation $s^-$. Let us now define the *fragility drift*:

$$V'_K(X, f_\lambda, K, s^-) = \frac{\partial^2 \xi}{\partial K \partial s^-}(K, s^-)$$

In practice, fragility always occurs as the result of *fragility*, indeed, by definition, we know that $\xi(\Omega, s^-) = s^-$, hence $V(X, f_\lambda, \Omega, s^-) = 1$. The *fragility drift* measures the speed at which fragility departs from its original value 1 when $K$ departs from the center $\Omega$.

## Second-order Fragility

The *second-order fragility* is the second order derivative of the tail estimate $\xi$ with respect to the semi-absolute deviation $s^-$:

$$V'_{s^-}(X, f_\lambda, K, s^-) = \frac{\partial^2 \xi}{\left(\partial s^-\right)^2}(K, s^-)$$

As we shall see later, the *second-order fragility* drives the bias in the estimation of stress tests when the value of $s^-$ is subject to uncertainty, through Jensen inequality.

# Definitions of Robustness and Antifragility

Antifragility is not the simple opposite of fragility, as we saw in Table 1. Measuring antifragility, on the one hand, consists of the flipside of fragility on the right-hand side, but on the other hand requires a control on the *robustness* of the probability distribution on the left-hand side. From that aspect, unlike fragility, antifragility cannot be summarized in one single figure but necessitates at least two of them.

When a random variable depends on another source of randomness: $Y_\lambda = \varphi(X_\lambda)$, we shall study the antifragility of $Y_\lambda$ with respect to that of $X_\lambda$ and to the properties of the function $\varphi$.

### DEFINITION OF ROBUSTNESS

Let $(X_\lambda)$ be a one-parameter family of random variables with pdf $f_\lambda$. Robustness is an upper control on the *fragility* of $X$, which resides on the left hand side of the distribution.

*We say that $f_\lambda$ is b-robust beyond stress level $K < \Omega$ if $V(X_\lambda, f_\lambda, K', s(\lambda)) \le b$ for any $K' \le K$. In other words, the <u>robustness</u> of $f_\lambda$ on the half-line ($-\infty, K$] is* $R_{(-\infty, K]}(X_\lambda, f_\lambda, K, s^-(\lambda)) = \max_{K' \le K} V(X_\lambda, f_\lambda, K', s^-(\lambda))$, *so that b-robustness simply means* $R_{(-\infty, K]}(X_\lambda, f_\lambda, K, s^-(\lambda)) \le b$.

We also define *b-robustness over a given interval* $[K_1, K_2]$ by the same inequality being valid for any $K' \in [K_1, K_2]$. In this case we use $R_{[K_1, K_2]}(X_\lambda, f_\lambda, K, s^-(\lambda)) = \max_{K_1 \le K' \le K_2} V(X_\lambda, f_\lambda, K', s^-(\lambda))$.

Note that the *lower R*, the tighter the control and the *more robust* the distribution $f_\lambda$.

Once again, the definition of *b-robustness* can be transposed, using finite differences $V(X_\lambda, f_\lambda, K', s^-(\lambda), \Delta s)$.

In practical situations, setting a material upper bound $b$ to the fragility is particularly important: one need to be able to come with actual estimates of the impact of the error on the estimate of the left-semi-deviation. However, when dealing with certain class of models, such as Gaussian, exponential of stable distributions, we may be lead to consider asymptotic definitions of robustness, related to certain classes.

For instance, for a given decay exponent $a > 0$, assuming that $f_\lambda(x) = O(e^{ax})$ when $x \to -\infty$, the *a*-exponential asymptotic robustness of $X_\lambda$ below the level $K$ is:

$$R_{\exp}(X_\lambda, f_\lambda, K, s^-(\lambda), a) = \max_{K' \le K} \left( e^{a(\Omega - K')} V(X_\lambda, f_\lambda, K', s^-(\lambda)) \right)$$

If one of the two quantities $e^{a(\Omega - K')} f_\lambda(K')$ or $e^{a(\Omega - K')} V(X_\lambda, f_\lambda, K', s^-(\lambda))$ is not bounded from above when $K' \to -\infty$, then $R_{\exp} = +\infty$ and $X_\lambda$ is considered as not *a*-exponentially robust.

Similarly, for a given power $\alpha > 0$, and assuming that $f_\lambda(x) = O(x^{-\alpha})$ when $x \to -\infty$, the *α*-power asymptotic robustness of $X_\lambda$ below the level $K$ is:

$$R_{\text{pow}}(X_\lambda, f_\lambda, K, s^-(\lambda), a) = \max_{K' \le K} \left( (\Omega - K')^{\alpha - 2} V(X_\lambda, f_\lambda, K', s^-(\lambda)) \right)$$

If one of the two quantities $(\Omega - K')^\alpha f_\lambda(K')$ or $(\Omega - K')^{\alpha-2} V(X_\lambda, f_\lambda, K', s^-(\lambda))$ is not bounded from above when $K' \quad -\infty$, then $R_{pow} = +\infty$ and $X_\lambda$ is considered as not $\alpha$-power robust. Note the exponent $\alpha - 2$ used with the fragility, for homogeneity reasons, e.g. in the case of stable distributions.

When a random variable $Y_\lambda = \varphi(X_\lambda)$ depends on another source of risk $X_\lambda$.

**Definition 2a, Left-Robustness (monomodal distribution)**. *A payoff $y = \varphi(x)$ is said (a,b)-robust below $L = \varphi(K)$ for a source of randomness $X$ with pdf $f_\lambda$ assumed monomodal if, letting $g_\lambda$ be the pdf of $Y = \varphi(X)$, one has, for any $K' \le K$ and $L' = \varphi(K')$ :*

$$V_X\left(Y, g_\lambda, L', s^-(\lambda)\right) \le aV\left(X, f_\lambda, K', s^-(\lambda)\right) + b \tag{4}$$

The quantity $b$ is of order deemed of "negligible utility" (subjectively), that is, does not exceed some tolerance level in relation with the context, while $a$ is a scaling parameter between variables $X$ and $Y$.

Note that robustness is in effect impervious to changes of probability distributions. Also note that this measure robustness ignores first order variations since owing to their higher frequency, these are detected (and remedied) very early on.

**Example of Robustness (Barbells):**

   **a.** trial and error with bounded error and open payoff

   **b.** for a "barbell portfolio" with allocation to numeraire securities up to 80% of portfolio, no perturbation below $K$ set at 0.8 of valuation will represent any difference in result, i.e. $q = 0$. The same for an insured house (assuming the risk of the insurance company is not a source of variation), no perturbation for the value below $K$, equal to minus the insurance deductible, will result in significant changes.

   **c.** a bet of amount B (limited liability) is robust, as it does not have any sensitivity to perturbations below 0.

### *DEFINITION OF ANTIFRAGILITY*

The second condition of *antifragility* regards the *right hand side* of the distribution. Let us define the *right-semi-deviation* of $X$ :

$$s^+(\lambda) = \int_\Omega^{+\infty} (x - \Omega) f_\lambda(x) dx$$

And, for $H > L > \Omega$ :

$$\xi^+(L, H, s^+(\lambda)) = \int_L^H (x - \Omega) f_\lambda(x) dx$$

$$W(X, f_\lambda, L, H, s^+) = \frac{\partial \xi^+(L, H, s^+)}{\partial s^+} = \left( \int_L^H (x - \Omega) \frac{\partial f_\lambda}{\partial \lambda}(x) dx \right) \left( \int_\Omega^{+\infty} (x - \Omega) \frac{\partial f_\lambda}{\partial \lambda}(x) dx \right)^{-1}$$

When $Y = \varphi(X)$ is a variable depending on a source of noise $X$, we define:

$$W_X(Y, g_\lambda, \varphi(L), \varphi(H), s^+) = \left( \int_{\varphi(L)}^{\varphi(H)} (y - \varphi(\Omega)) \frac{\partial g_\lambda}{\partial \lambda}(y) dy \right) \left( \int_\Omega^{+\infty} (x - \Omega) \frac{\partial f_\lambda}{\partial \lambda}(x) dx \right)^{-1}$$

**Definition 2b, Antifragility (monomodal distribution)**. *A payoff $y = \varphi(x)$ is locally antifragile over the range $[L, H]$ if*

   **1.** *It is b-robust below $\Omega$ for some $b > 0$*

   **2.** $W_X\left(Y, g_\lambda, \varphi(L), \varphi(H), s^+(\lambda)\right) \ge aW\left(X, f_\lambda, L, H, s^+(\lambda)\right)$ *where* $a = \dfrac{u^+(\lambda)}{s^+(\lambda)}$

The scaling constant $a$ provides homogeneity in the case where the relation between $X$ and $y$ is linear. In particular, nonlinearity in the relation between $X$ and $Y$ impacts robustness.

The second condition can be replaced with finite differences $\Delta u$ and $\Delta s$, as long as $\Delta u/u = \Delta s/s$.

REMARKS

**Fragility is *K*-specific**. We are only concerned with adverse events below a certain pre-specified level, the breaking point. Exposures A can be more fragile than exposure B for $K = 0$, and much less fragile if $K$ is, say, 4 mean deviations below 0. We may need to use finite $\Delta s$ to avoid situations as we will see of vega-neutrality coupled with short left tail.

**Effect of using the wrong distribution *f***: Comparing $V(X, f, K, s^-, \Delta s)$ and the alternative distribution $V(X, f^*, K, s^*, \Delta s)$, where $f^*$ is the "true" distribution, the measure of fragility provides an acceptable indication of the sensitivity of a given outcome – such as a risk measure – to model error, provided no "paradoxical effects" perturb the situation. Such "paradoxical effects" are, for instance, a change in the direction in which certain distribution percentiles react to model parameters, like $s^-$. It is indeed possible that nonlinearity appears between the core part of the distribution and the tails such that when $s^-$ increases, the left tail starts fattening – giving a large measured fragility – then steps back – implying that the real fragility is lower than the measured one. The opposite may also happen, implying a dangerous under-estimate of the fragility. These nonlinear effects can stay under control provided one makes some regularity assumptions on the <u>actual</u> distribution, as well as on the measured one. For instance, paradoxical effects are typically avoided under at least one of the following three hypotheses:

    **a.**    The class of distributions in which both $f$ and $f^*$ are picked all are monomodal, with monotonous dependence of percentiles with respect to one another.

    **b.**    The difference between percentiles of $f$ and $f^*$ has constant sign (i.e. $f^*$ is either *always* wider or *always* narrower than $f$ at any given percentile)

    **c.**    For any strike level $K$ (in the range that matters), the fragility measure $V$ monotonously depends on $s^-$ on the whole range where the true value $s^*$ can be expected. This is in particular the case when partial derivatives $\partial^k V / \partial s^k$ all have the same sign at measured $s^-$ up to some order $n$, at which the partial derivative has that same constant sign over the whole range on which the true value $s^*$ can be expected. This condition can be replaced by an assumption on finite differences approximating the higher order partial derivatives, where $n$ is large enough so that the interval $[s^- \pm n\Delta s]$ covers the range of possible values of $s^*$. Indeed, in this case, the finite difference estimate of fragility uses evaluations of $\xi$ at points spanning this interval.

**Unconditionality of the shortfall measure $\xi$** : Many, when presenting shortfall, deal with the conditional shortfall $\int_{-\infty}^{K} x\ f(x)\ dx \Big/ \int_{-\infty}^{K} f(x)\ dx$; while such measure might be useful in some circumstances, its sensitivity is not indicative of fragility in the sense used in this discussion. The unconditional tail expectation $\xi = \int_{-\infty}^{K} x f(x)\ dx$ is more indicative of exposure to fragility. It is also preferred to the raw probability of falling below $K$, which is $\int_{-\infty}^{K} f(x)\ dx$, as the latter does not include the consequences. For instance, two such measures $\int_{-\infty}^{K} f(x)\ dx$ and $\int_{-\infty}^{K} g(x)\ dx$ may be equal over broad values of $K$; but the expectation $\int_{-\infty}^{K} x f(x)\ dx$ can be much more consequential than $\int_{-\infty}^{K} x g(x)\ dx$ as the cost of the break can be more severe and we are interested in its "vega" equivalent.

# Applications to Model Error

In the cases where Y depends on X, among other variables, often x is treated as non-stochastic, and the underestimation of the volatility of x maps immediately into the underestimation of the left tail of Y under two conditions:

    a- X is stochastic and its stochastic character is ignored (as if it had zero variance or mean deviation)

    b- Y is concave with respect to X in the negative part of the distribution, below $\Omega$

**"Convexity Bias" or effect from Jensen's Inequality**: Further, missing the stochasticity under the two conditions a) and b) , in the event of the concavity applying above $\Omega$ leads to the negative convexity bias from the lowering effect on the expectation of the dependent variable Y.

### CASE 1- APPLICATION TO DEFICITS

**Example:** A government estimates unemployment for the next three years as averaging 9%; it uses its econometric models to issue a forecast balance $B$ of 200 billion deficit in the local currency. But it misses (like almost everything in economics) that unemployment is a stochastic variable. Employment over 3 years periods has fluctuated by 1% on average. We can calculate the effect of the error with the following:

- Unemployment at 8% , Balance B(8%) = -75 bn (improvement of 125bn)

- Unemployment at 9%, Balance B(9%)= -200 bn

- Unemployment at 10%, Balance B(10%)= --550 bn (worsening of 350bn)

The convexity bias from underestimation of the deficit is by -112.5bn, since $\frac{B(8\%)+B(10\%)}{2}=-312.5$

Further look at the probability distribution caused by the missed variable (assuming to simplify deficit is Gaussian with a Mean Deviation of 1% )

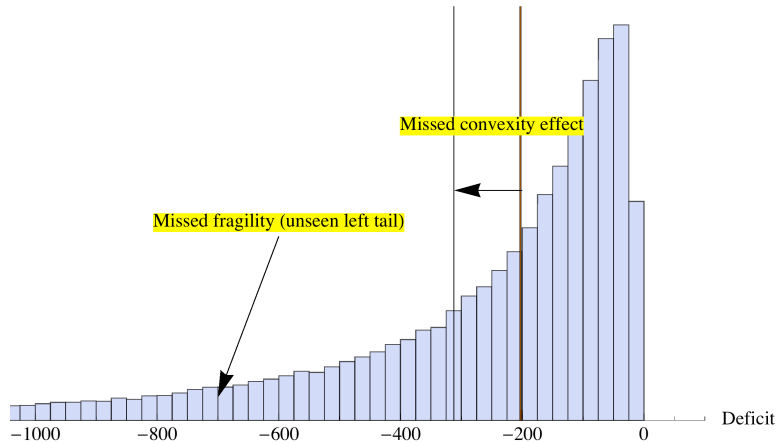

*Figure 6* **CONVEXITY EFFECTS ALLOW THE DETECTION OF BOTH MODEL BIAS AND FRAGILITY.** *Illustration of the example; histogram from Monte Carlo simulation of government deficit as a left-tailed random variable simply as a result of randomizing unemployment of which it is a convex function. The method of point estimate would assume a Dirac stick at -200, thus underestimating both the **expected** deficit (-312) and the skewness (i.e., fragility) of it.*

**Adding Model Error and Metadistributions**: Model error should be integrated in the distribution as a stochasticization of parameters. *f* and *g* should subsume the distribution of all possible factors affecting the final outcome (including the metadistribution of each). The so-called "perturbation" is not necessarily a change in the parameter so much as it is a means to verify whether *f* and *g* capture the full shape of the final probability distribution.

Any situation with a bounded payoff function that organically truncates the left tail at *K* will be impervious to all perturbations affecting the probability distribution below *K*.

For *K* = 0, the measure equates to mean negative semi-deviation (more potent than negative semi-variance or negative semi-standard deviation often used in financial analyses).

### MODEL ERROR AND SEMI-BIAS AS NONLINEARITY FROM MISSED STOCHASTICITY OF VARIABLES

Model error often comes from missing the existence of a random variable that is significant in determining the outcome (say option pricing without credit risk). We cannot detect it using the heuristic presented in this paper but as mentioned earlier the error goes in the opposite direction as model tend to be richer, not poorer, from overfitting. But we can detect the model error from missing the stochasticity of a variable or underestimating its stochastic character (say option pricing with non-stochastic interest rates or ignoring that the "volatility" σ can vary).

**Missing Effects**: The study of model error is not to question whether a model is precise or not, whether or not it tracks reality; it is to ascertain the first and second order effect from missing the variable, insuring that the errors from the model don't have missing higher order terms that cause severe unexpected (and unseen) biases in one direction because of convexity or concavity, in other words, whether or not the model error causes a change in z.

### Model Bias, Second Order Effects, and Fragility

Having the right model (which is a very generous assumption), but being uncertain about the parameters will invariably lead to an increase in model error in the presence of convexity and nonlinearities.

As a generalization of the deficit/employment example used in the previous section, say we are using a simple function:

$$f\left(x\,|\,\overline{\alpha}\right) \tag{5}$$

Where $\overline{\alpha}$ is supposed to be the average expected rate, where we take $\varphi$ as the distribution of $\alpha$ over its domain $\wp_\alpha$

$$\overline{\alpha} = \int_{\wp_\alpha} \alpha\,\varphi(\alpha)\,d\alpha \tag{6}$$

The mere fact that $\alpha$ is uncertain (since it is estimated) might lead to a bias if we perturb from the <u>outside</u> (of the integral), i.e. stochasticize the parameter deemed fixed. Accordingly, the convexity bias is easily measured as the difference between a) $f$ integrated across values of potential $\alpha$ and b) $f$ estimated for a single value of $\alpha$ deemed to be its average. The convexity bias $\omega_A$ becomes:

$$\omega_A \equiv \int_{\wp_x} \int_{\wp_\alpha} f\left(x\,|\,\alpha\right)\varphi\left(\alpha\right)d\alpha\,dx - \int_{\wp_x} f\left(x\,\middle|\,\left(\int_{\wp_\alpha} \alpha\,\varphi\left(\alpha\right)d\alpha\right)\right)dx \tag{7}$$

And $\omega_B$ the missed fragility is assessed by comparing the two integrals below $K$, in order to capture the effect on the left tail:

$$\omega_B(K) \equiv \int_{-\infty}^{K} \int_{\wp_\alpha} f\left(x\,|\,\alpha\right)\varphi\left(\alpha\right)d\alpha\,dx - \int_{-\infty}^{K} f\left(x\,\middle|\,\left(\int_{\wp_\alpha} \alpha\,\varphi\left(\alpha\right)d\alpha\right)\right)dx \tag{8}$$

Which can be approximated by an interpolated estimate obtained with two values of $\alpha$ separated from a mid point by $\Delta\alpha$ a mean deviation of $\alpha$ and estimating

$$\omega_B(K) \equiv \int_{-\infty}^{K} \frac{1}{2}\left(f\left(x\,|\,\overline{\alpha}+\Delta\alpha\right)+f\left(x\,|\,\overline{\alpha}-\Delta\alpha\right)\right)dx - \int_{-\infty}^{K} f(x\,|\,\overline{\alpha})\,dx \tag{8}$$

We can probe $\omega_B$ by point estimates of $f$ at a level of $X \leq K$

$$\omega_B'(X) = \frac{1}{2}\left(f\left(X\,|\,\overline{\alpha}+\Delta\alpha\right)+f\left(X\,|\,\overline{\alpha}-\Delta\alpha\right)\right) - f(X\,|\,\overline{\alpha}) \tag{9}$$

So that

$$\omega_B(K) = \int_{-\infty}^{K} \omega_B'(x)\,dx$$

which leads us to the fragility heuristic. In particular, if we assume that $\omega_B'(X)$ has a constant sign for $X \leq K$, then $\omega_B(K)$ has the same sign.

# The Fragility/Model Error Detection Heuristic (detecting $\omega_A$ and $\omega_B$ when cogent)

*Example 1 (Detecting Tail Risk Not Shown By Stress Test, $\omega_B$). The famous firm Dexia went into financial distress a few days after passing a stress test "with flying colors".*

*If a bank issues a so-called "stress test" (something that has never worked in history), off a parameter (say stock market) at -15%. We ask them to recompute at -10% and -20%. Should the exposure show negative asymmetry (worse at -20% than it improves at -10%), we deem that their risk increases in the tails. There are certainly hidden tail exposures and a definite higher probability of blowup in addition to exposure to model error.*

*Note that it is somewhat more effective to use our measure of shortfall in Definition, but the method here is effective enough to show hidden risks, particularly at wider increases (try 25% and 30% and see if exposure shows increase). Most effective would be to use power-law*

*distributions and perturb the tail exponent to see symmetry.*

**Example 2 (Detecting Tail Risk in Overoptimized System, $\omega_B$).** *Raise airport traffic 10%, lower 10%, take average expected traveling time from each, and check the asymmetry for nonlinearity. If asymmetry is significant, then declare the system as overoptimized. (Both $\omega_A$ and $\omega_B$ as thus shown.*

**Example 3 (Detecting error in a probability framework, $\omega_B$).** Change parameter up by a mean deviation, compute left integral below K; lower it by a mean deviation; compute and compare. If there is asymmetry, then the situation is precarious and model error is large to the extent of the asymmetry.

The same procedure uncovers both fragility and consequence of model error (potential harm from having wrong probability distribution, a thin-tailed rather than a fat-tailed one). For traders (and see Gigerenzer's discussions, Gigerenzer and Brighton (2009), Gigerenzer and Goldstein (1996)) playing with second order effects of simplistic tools can be more effective than more complicated and harder to calibrate methods. See also the intuition of fast and frugal in Derman and Wilmott (2009), Haug and Taleb (2011).

*The Heuristic*:

Taleb, Canetti et al (2012) developed a heuristic in parallel with this paper to apply to stress testing or more generally, valuations.

*1- First Step (first order). Take $\psi$ a valuation. Measure the sensitivity to all parameters p determining V over finite ranges $\Delta p$. If materially significant, check if stochasticity of parameter is taken into account by risk assessment. If not, then stop and declare the risk as grossly mismeasured (no need for further risk assessment). (Note that Ricardo's wine-cloth example miserably fails the first step upon stochasticizing either.)*

*2-Second Step (second order). For all parameters p compute the second order* $H(\Delta p) \equiv \frac{\mu'}{\mu}$, *where* $\mu'(\Delta p) \equiv \frac{1}{2}\left( f\left( p + \frac{1}{2}\Delta p \right) + f\left( p - \frac{1}{2}\Delta p \right) \right)$

*2-Third Step. Note parameters for which H is significantly > or < 1*

*Properties of the Heuristic*:

*i- **Fragility**: V' is a more accurate indicator of fragility than V over $\Delta p$ when p is stochastic or subjected to estimation errors with mean deviation $\Delta p$*

*ii- **Model Error**: A model M(p) with parameter p held constant underestimates the fragility of payoff from x under perturbation $\Delta p$ if H>1,*
*iii- if H=1, the exposure to x is robust over $\Delta p$ and model error over p is inconsequential.*
*iv- if H remains $\geq 1$ for larger and larger $\Delta p$, then the heuristic is broad (absence of pseudo-convexities)*

We can apply the method to V in Equation 1, as it becomes a perturbation of a perturbation, (in *Dynamic Hedging* "vvol", or "volatility of volatility" or in later lingo vvol), $H = \frac{V\left( x, f, K, \Delta s + \frac{\Delta s}{2} \right) + V\left( x, f, K, \Delta s - \frac{\Delta s}{2} \right)}{2V(x, f, K, \Delta s)}$ where K is the fragility threshold, x is a random variable describing outcomes, $\Delta p$ is a set perturbation and f the probability measure used to compute $\zeta$.

Note that for K set at $\infty$, the heuristic becomes a simple detection of model bias from the effect of Jensen's inequality when stochasticizing a term held to be deterministic, the measure of $\omega_A$.

The heuristic has the ability to "fill-in the tail", by extending further down into the probability distribution as $\Delta p$ increases. It is best to perturb the tail exponent of a power law.

Remarks:

**a.** Simple heuristics have a robustness (in spite of a possible bias) compared to optimized and calibrated measures. Ironically, it is from the multiplication of convexity biases and the potential errors from missing them that calibrated models that work in-sample underperform heuristics out of sample.
**b.** It allows to detection of the effect of the use of the wrong probability distribution without changing probability distribution (just from the dependence on parameters).
**c.** It outperforms all other commonly used measures of risk, such as CVaR, "expected shortfall", stress-testing, and similar methods have been proven to be completely ineffective.
**d.** It does not require parameterization beyond varying $\Delta p$.

# Comparison of the Heuristic to Other Methods

**CVaR & VaR**: these are totally ineffective, no need for further discussion here (or elsewhere) as they have been shown to be so empirically and mathematically. But perturbation can reveal convexities and concavities.

**Stress Testing**. One of the authors has shown where these can be as ineffective owing to risk hiding in the tail below the stress test. See Taleb (2009) on why the level K of the stress test is arbitrary and cannot be appropriately revealed by the past realizations of the random variable. But if stress tests show an increase in risk at lower and lower levels of stress, then the position reveals exposure in the tails.

Note that hidden risks reside in the tails as they are easy to hide there, undetected by conventional methods and tend to hide there.

# Further Applications

In parallel works, applying the *"simple heuristic"* allows us to detect the following "hidden short options" problems by merely perturbating a certain parameter $p$:

   **a.** Size and pseudo-economies of scale.

     **i.** size and squeezability (nonlinearities of squeezes in costs per unit)

   **b.** Specialization (Ricardo) and variants of globalization.

     **i.** missing stochasticity of variables (price of wine).

     **ii.** specialization and nature.

   **c.** Portfolio optimization (Markowitz)

   **d.** Debt

   **e.** Budget Deficits: convexity effects explain why uncertainty lengthens, doesn't shorten expected deficits.

   **f.** Iatrogenics (medical) or how some treatments are concave to benefits, convex to errors.

   **g.** Disturbing natural systems

   **h.** It detects fragility to forecasting errors in projection as these reside in convexity of duration/costs to uncertainty

   **i.** Hidden tail exposure from dependence on a source of energy, etc. ("squeezability argument")

   j. Medical applications to assess relative iatrogenics from the nonlinearity in response.

## Acknowledgments

Bruno Dupire, Emanuel Derman, Jean-Philippe Bouchaud, Elie Canetti.

JP Morgan, New York, June 16, 2011; CFM, Paris, June 17, 2011; GAIM Conference, Monaco, June 21, 2011; Max Planck Institute, BERLIN, Summer Institute on Bounded Rationality 2011 - *Foundations of an Interdisciplinary Decision Theory*- June 23, 2011; Eighth International Conference on Complex Systems - BOSTON, July 1, 2011, Columbia University September 24 2011.